# On Financial Markets Based on Telegraph Processes

Nikita Ratanov† and Alexander Melnikov∗‡

† Universidad del Rosario, Bogotá, Colombia

‡ University of Alberta, Edmonton, Canada

The paper develops a new class of financial market models. These models are based on generalized telegraph processes: Markov random flows with alternating velocities and jumps occurring when the velocities are switching. While such markets may admit an arbitrage opportunity, the model under consideration is arbitrage-free and complete if directions of jumps in stock prices are in a certain correspondence with their velocity and interest rate behaviour. An analog of the Black-Scholes fundamental differential equation is derived, but, in contrast with the Black-Scholes model, this equation is hyperbolic. Explicit formulas for prices of European options are obtained using perfect and quantile hedging.

## 1. Introduction

Since the 1970s, the mathematics of finance has been developing in the direction of creating progressively more general models for price evolutions of basic financial assets, from Brownian motion to the almost all-encompassing semimartingale processes. This has led to the creation of theory of no-arbitrage and completeness, as well as hedging and optimal investment (see Merton [18], Duffie [9], Delbaen and Schachermayer [8], Karatzas and Shreve [16], etc). Although it would be difficult to improve these theoretical findings in terms of structural generality, the efforts to calculate exact theoretically and practically significant formulas for option pricing and corresponding optimal investment have been successful only for those models of financial markets in which the increments of underlying random processes are independent (Wiener, Poisson, Lévy processes etc.).

Meanwhile, the development of non-semimartingale models focused mainly on accounting for the dependence of asset prices on the past (long-term memory processes, fractional Brownian motion etc.). However, as of today there is still no commonly accepted theory on this topic, nor adequate uses of existing theoretical results in practice (see, for instance, Björk and Hult [4]).

Another direction, which lies somewhat in between the two trends mentioned above, utilizes markovian dependence on the past and the technique of Markov random processes. In our opinion, this approach and the corresponding methods (see Elliott and van der Hoek [10]) are still inadequately reflected in contemporary financial mathematics.

Our paper deals mainly with this direction of study of financial markets. As a basis for building the model, we take a Markov process $\sigma(t)$ with values $\pm 1$ and transition probability intensities $\lambda_\pm$. Using these, we define processes $c_{\sigma(t)} = c_\pm$, $h_{\sigma(t)} = h_\pm$ and $r_{\sigma(t)} = r_\pm$, $r_\pm > 0$. Let us introduce $X^\sigma(t) = \int_0^t c_{\sigma(s)} ds$ and a pure jump process $J^\sigma = J^\sigma(t)$ with alternating jumps of sizes $h_\pm$. The evolution of the risky asset $S(t)$ is determined by the stochastic exponent of the sum $X^\sigma + J^\sigma$. The risk-free asset is given by the usual exponent of the process $Y^\sigma = Y^\sigma(t) = \int_0^t r_{\sigma(s)} ds$. Here and below the superscript $\sigma$ indicates the starting value $\sigma = \sigma(0)$ of $\sigma(t)$.

---

∗Corresponding author. Email: melnikov@ualberta.ca

In view of such trajectories, the market is set up as a continuous process that evolves with velocity $c_+$ or $c_-$, changes the direction of movement from $c_\pm$ to $c_\mp$ and exhibits jumps of size $h_\pm$ whenever velocity changes.

The interest rate in the market is stochastic with the values $r_\pm$ such that $(c_\pm - r_\pm)h_\pm < 0$ which means that the current trend of discounted prices and the direction of the next price jump should be opposite. A process $X^\sigma = X^\sigma(t)$ is defined by a pair of states $(c_\pm, \lambda_\pm)$ and is called a *telegraph process with states* $(c_\pm, \lambda_\pm)$. It describes continuous price trends (upward or downward) between random instants. Changes in these trends are accompanied by jumps of sizes $h_\pm$.

Our model uses parameters $c_\pm$ to capture bullish and bearish trends in a market evolution, and values $h_\pm$ to describe sizes of possible crashes, jumps and spikes. Thus, we study a model that is both realistic and general enough to enable us to incorporate different trends and extreme events. At the same time, as it will be seen in further sections, the model allows us to get closed form solutions for hedging and investment problems.

Sections 2-3 deal with the properties of such processes and the mathematical model of the market. Among the relevant results, we construct a unique martingale measure based on Girsanov's theorem. This measure guarantees the absence of arbitrage in our setting and shows that, under some scaling normalization, our model converges to that of Black-Scholes in distribution. In the next section, devoted to perfect hedging of options, we derive the fundamental equation, which, unlike the classical Black-Scholes model, is hyperbolic. In Section 5, we calculate the price of a call option and its simplifications including the case of Merton's model. In section 6, we study the questions relevant to imperfect hedging in the context of the considered model.

Telegraph processes have been studied before in different probabilistic aspects (see, for instance, Goldstein [13], Kac [15], Orsingher [19] and Zacks [24]). These processes have been exploited for stochastic volatility modelling (Di Masi et al [7], Melnikov et al [17]) as well as for obtaining a "telegraph analog" of the Black-Scholes model (Di Crescenzo and Pellerey [6]). In contrast with the paper by Di Crescenzo and Pellerey, we use a more complicated and delicate construction of such a model to avoid arbitrage and to develop an adequate option pricing theory in this framework.

## 2. Telegraph processes and auxiliary results

Let $(\Omega, \mathcal{F}, \mathsf{P})$ be a complete probability space and $\sigma = \sigma(t)$, $t \geq 0$ be a right-continuous Markov process, taking values $\pm 1$ and having intensities $\lambda_\pm > 0$:

$$\mathsf{P}(\sigma(t + \Delta t) = +1 \mid \sigma(t) = -1) = \lambda_- \Delta t + o(\Delta t), \tag{1}$$

$$\mathsf{P}(\sigma(t + \Delta t) = -1 \mid \sigma(t) = +1) = \lambda_+ \Delta t + o(\Delta t), \tag{2}$$

as $\Delta t \to 0$. The initial state $\sigma = \sigma(0)$ of the process $\sigma(t), t \geq 0$ is deterministic and equal to $+1$ or $-1$.

Further, we will consider all processes adapted to the filtration $\mathbf{F} = (\mathbf{F}_t^\sigma)_{t \geq 0}$ ($\mathbf{F}_0^\sigma = \{\emptyset, \Omega\}$), generated by $\sigma(t)$, $t \geq 0$, starting at $\sigma$, $\sigma = \sigma(0) = \pm 1$. We suppose that the filtration satisfies the "usual conditions". Recall (see e.g. Karatzas and Shreve [16]) that a filtration $(\mathbf{F}_t)$ is said to satisfy the usual conditions if it is right-continuous and $(\mathbf{F}_0)$ contains all $\mathsf{P}$-negligible sets of $\mathbf{F}$. Under this assumption, the process $\sigma(t)$, $t \geq 0$ can be viewed as a Markov flow controlled by transition probabilities (2.1)-(2.2). The time intervals $\tau_j - \tau_{j-1}$, $j = 1, 2, \ldots$ ($\tau_0 = 0$), separated by instants of value changes $\tau_j = \tau_j^\sigma$, $j = 1, 2, \ldots$ are independent. Also, we denote $\mathsf{P}_\sigma$ the conditional probability with respect to the initial state $\sigma$, and $\mathsf{E}_\sigma$ the expectation with respect to $\mathsf{P}_\sigma$.

Let $N^\sigma(t)$ be the number of switches on $[0, t]$ of the process $\sigma(t)$, $t \geq 0$ starting at $\sigma$, $\sigma = \sigma(0) = \pm 1$. Note (see, for instance, Ross [23]) that $N^\sigma = N^\sigma(t)$, $t \geq 0$ is a Poisson process with alternating intensities $\lambda_\sigma, \lambda_{-\sigma}, \lambda_\sigma, \ldots$, $\sigma = \pm 1$.

For given numbers $c_- \leq c_+$ and $h_\pm$ we define the following processes

$$X^\sigma(t) = \int_0^t c_{\sigma(s)} ds \tag{3}$$

and

$$J^\sigma(t) = \sum_{j=1}^{N^\sigma(t)} h_{\sigma(\tau_j-)}, \quad t \geq 0, \tag{4}$$

where $\sigma(\tau_j-)$ is the left-limit of $\sigma(t)$ at $\tau_j$. As before, the superscript $\sigma = \sigma(0)$ indicates the starting condition of the process $\sigma(t)$, $t \geq 0$.

**Definition 2.1:** The processes $X^\sigma$, $\sigma = \pm 1$ are called (inhomogeneous) telegraph processes with states $(c_-, \lambda_-)$, $(c_+, \lambda_+)$ with starting at $\sigma$.

For $\lambda_- = \lambda_+$ and $-c_- = c_+ = c$, the process $X^\sigma = \sigma c \int_0^t (-1)^{N(s)} ds$, $t \geq 0$ is usually referred to as (integrated) telegraph process (see Goldstein [13] and Kac [14]).

The following lemma is evident.

**Lemma 2.2:** Let the processes $X^\sigma = X^\sigma(t)$ and $\tilde{X}^\sigma = \tilde{X}^\sigma(t)$, $t \geq 0$ be telegraph processes with states $(c_\pm, \lambda_\pm)$ and $(\tilde{c}_\pm, \lambda_\pm)$ respectively, governed by the common Markov process $\sigma = \sigma(t)$. Then they are linearly connected $\tilde{X}^\sigma(t) = aX^\sigma(t) + bt$, $t \geq 0$, where

$$a = a_{\tilde{c}} = \frac{\tilde{c}_+ - \tilde{c}_-}{c_+ - c_-}, \quad b = b_{\tilde{c}} = \frac{c_+ \tilde{c}_- - c_- \tilde{c}_+}{c_+ - c_-}, \tag{5}$$

and $ac_- + b = \tilde{c}_-$, $ac_+ + b = \tilde{c}_+$.

The next theorem could be considered as a version of the Doob-Meyer decomposition for telegraph processes with alternating intensities.

**Theorem 2.3:** Let $X^\sigma$ be the telegraph process with states $(c_-, \lambda_-)$ and $(c_+, \lambda_+)$, defined in (3), and $J^\sigma$ be the jump process, defined in (4), $\sigma = \pm 1$. Then $X^\sigma + J^\sigma$ is a martingale if and only if

$$\lambda_- h_- = -c_-, \qquad \lambda_+ h_+ = -c_+. \tag{6}$$

**Remark 1:** In particular, any (nontrivial) telegraph process without jumps (i.e. if $h_\pm = 0$) never possess the martingale measure.

**Proof:** In the particular case $\lambda_\pm = \lambda$, $h_\pm = h$, $c_\pm = c$, the theorem follows from the martingale property of $N(t) - \lambda t$, $t \geq 0$. In a general situation, we have

$$\mathsf{E}(J^\sigma(t) \mid \mathbf{F}_s^\sigma) = J^\sigma(s) + \gamma H(t-s) + \lambda_{\sigma(s)} a_{\sigma(s)} \frac{1 - e^{-\Lambda(t-s)}}{\Lambda}, \tag{7}$$

$$\mathsf{E}(X^\sigma(t) \mid \mathbf{F}_s^\sigma) = X^\sigma(s) + g(t-s) + \lambda_{\sigma(s)} d_{\sigma(s)} \frac{1 - e^{-\Lambda(t-s)}}{\Lambda}, \quad \sigma = \pm 1. \tag{8}$$

Here $H = h_- + h_+$, $\Lambda = \lambda_- + \lambda_+$, $\gamma = \frac{\lambda_- \lambda_+}{\Lambda}$, $g = \frac{c_+ \lambda_- + c_- \lambda_+}{\Lambda}$, and $a_\sigma = \frac{\lambda_\sigma h_\sigma - \lambda_{-\sigma} h_{-\sigma}}{\Lambda}$, $d_\sigma = \frac{c_\sigma - c_{-\sigma}}{\Lambda}$, $\sigma = \pm 1$.

To ensure the martingale property, we have to solve the equations

$$\begin{cases} \gamma H + g = 0, \\ a_- + d_- = 0, \\ a_+ + d_+ = 0, \end{cases}$$

which clearly leads us to (6).                                                                                      □

Next, we study the properties of telegraph processes under a change of measure. Let $X_*^\sigma$ be the telegraph process with the states $(c_\pm^*, \lambda_\pm)$, and $J_*^\sigma = -\sum_{j=1}^{N^\sigma(t)} c_{\sigma(\tau_j-)}^*/\lambda_{\sigma(\tau_j-)}$ be the jump process with jump values $h_\pm^* = -c_\pm^*/\lambda_\pm > -1$. Consider a probability measure $\mathsf{P}_\sigma^*$ with a local density with respect to $\mathsf{P}_\sigma$:

$$Z^\sigma(t) = \frac{\mathrm{d}\mathsf{P}_\sigma^*}{\mathrm{d}\mathsf{P}_\sigma}\Big|_t = \mathcal{E}_t(X_*^\sigma + J_*^\sigma), \qquad 0 \leq t \leq T. \tag{9}$$

Here $\mathcal{E}_t(\cdot)$ denotes the stochastic exponential (see e. g. Melnikov et al [17]).

Using properties of stochastic exponentials, we obtain

$$Z^\sigma(t) = \mathrm{e}^{X_*^\sigma(t)} \kappa_*^\sigma(t), \tag{10}$$

where $\kappa_*^\sigma(t) = \prod_{s \leq t}(1 + \Delta J_*^\sigma(s))$ with $\Delta J_*^\sigma(s) = J_*^\sigma(s) - J_*^\sigma(s-)$.

The process $\kappa_*^\sigma = \kappa_*^\sigma(t)$ can be represented as $\kappa_*^\sigma(t) = \kappa_{N^\sigma(t),\sigma}^*$. Here $\sigma = \pm 1$ indicates the initial direction, and the sequence $\kappa_{n,\sigma}^*$ is defined as follows:

$$\kappa_{n,\sigma}^* = \kappa_{n-1,-\sigma}^*(1 + h_\sigma^*), \ n \geq 1, \qquad \kappa_{0,\sigma}^* \equiv 1. \tag{11}$$

Thus if $n = 2k$,

$$\kappa_{n,\sigma}^* = (1 + h_\sigma^*)^k(1 + h_{-\sigma}^*)^k,$$

and if $n = 2k + 1$,

$$\kappa_{n,\sigma}^* = (1 + h_\sigma^*)^{k+1}(1 + h_{-\sigma}^*)^k.$$

**Theorem 2.4 : (Girsanov theorem)** *Under the probability measure $\mathsf{P}_\sigma^*$,*
- *the process $N^\sigma = N^\sigma(t), \ t \geq 0$ is a Poisson process with intensities $\lambda_-^* = \lambda_- - c_-^* = \lambda_-(1 + h_-^*)$ and $\lambda_+^* = \lambda_+ - c_+^* = \lambda_+(1 + h_+^*)$.*
- *the process $X^\sigma = X^\sigma(t), \ 0 \leq t \leq T$ is a telegraph process with states $(c_-, \lambda_-^*)$ and $(c_+, \lambda_+^*)$.*

**Proof:** Let $\pi_n^{(\sigma)}(t) = \mathsf{P}_\sigma(N^\sigma(t) = n)$ and $\pi_{*,n}^{(\sigma)}(t) = \mathsf{P}_\sigma^*(N^\sigma(t) = n), \ n = 0, 1, 2 \ldots$, where $\sigma$ indicates the initial state. The probabilities $\pi_n^{(\sigma)}(t), \ \sigma = \pm 1$ solve the system

$$\begin{cases} \frac{\mathrm{d}\pi_n^{(\sigma)}}{\mathrm{d}t} = -\lambda_\sigma \pi_n^{(\sigma)}(t) + \lambda_\sigma \pi_{n-1}^{(-\sigma)}(t), \ t > 0 \\ \pi_n^{(\sigma)}\big|_{t=0} = 0, \ n \geq 1; \quad \pi_0^{(\sigma)}\big|_{t=0} = 1. \end{cases}$$

From (10)-(11), it follows that

$$\pi^{(\sigma)}_{*,n}(t) = \mathsf{E}_\sigma(Z^\sigma(t)\mathbf{1}_{\{N(t)=n\}}) = \kappa^*_{n,\sigma} \int_{-\infty}^{\infty} e^{ax+bt} p_n^{(\sigma)}(x,\,t)dx \qquad (12)$$

with $a$ and $b$ defined in (5), with $\tilde{c}_\pm$ replaced by $c^*_\pm$. Here $p_n^{(\sigma)}$, $n \geq 0$ are the (generalized) probability densities of the current position of the process $X^\sigma(t)$, which has $n$ turns with respect to the measure $\mathsf{P}_\sigma$. That is, for any borelian set $\Delta$

$$\mathsf{P}_\sigma(X^\sigma(t) \in \Delta,\, N^\sigma(t) = n) = \int_\Delta p_n^{(\sigma)}(x,\,t)dx. \qquad (13)$$

Note that densities $p_n^{(\sigma)}(x,\,t)$, $\sigma = \pm 1$ satisfy the equation

$$\frac{\partial p_n^{(\sigma)}}{\partial t} + c_\sigma \frac{\partial p_n^{(\sigma)}}{\partial x} = -\lambda_\sigma p_n^{(\sigma)} + \lambda_\sigma p_{n-1}^{(-\sigma)}, \qquad n \geq 1. \qquad (14)$$

To prove the above, we note that, conditioning on a switch at the time interval $(0,\,\Delta t)$,

$$p_n^{(\sigma)}(x,\,t+\Delta t) = (1 - \lambda_\sigma \Delta t)p_n^{(\sigma)}(x - c_\sigma \Delta t,\,t) + \lambda_\sigma \Delta t\, p_{n-1}^{(-\sigma)}(x,\,t) + o(\Delta t), \qquad \Delta t \to 0.$$

The system (14) should be supplied with zero initial conditions $p_n^{(\sigma)}\,|_{t=0}= 0$, $n \geq 1$ and $p_0^{(\sigma)} = e^{-\lambda_\sigma t}\delta(x - c_\sigma t)$.

Exploiting the equation (14), we obtain from (12)

$$\frac{d\pi^{(\sigma)}_{*,n}}{dt} = (b - \lambda_\sigma + ac_\sigma)\pi^{(\sigma)}_{*,n}(t) + \lambda_\sigma(1 - c^*_\sigma/\lambda_\sigma)\pi^{(-\sigma)}_{*,n-1}(t).$$

The following evident equalities complete the proof:

$$b - \lambda_\sigma + ac_\sigma = c^*_\sigma - \lambda_\sigma = -\lambda^*_\sigma,$$

$$p_n^{(\sigma)}\,|_{t=0} = 0,\ n \geq 1, \qquad p_0^{(\sigma)}\,|_{t=0} = \delta(x).$$

$\square$

**Remark 2:** System (14) could be solved exactly. Probability densities $p_n^{(\pm)}$, $n \geq 0$ have a view $p_n^{(\pm)} = e^{(-\lambda_\pm + \nu v_\pm)t - \nu x}q_n^{(\pm)}$, where $\nu = \Delta\lambda/\Delta c = (\lambda_= - \lambda_-)/(c_+ - c_-)$ and $q_n^{(\pm)}$ are defined as follows (see [22]): $q_0^{(\pm)} = \delta(x - c_\pm t)$ (and hence $p_0^{(\pm)} = e^{-\lambda_\pm t}\delta(x - c_\pm t)$),

$$q_{2n}^{(+)} = \frac{\lambda_+^n \lambda_-^n}{(\Delta c)^{2n}} \cdot \frac{(c_+ t - x)^{n-1}(x - c_- t)^n}{(n-1)!n!}\theta_t, \qquad q_{2n}^{(-)} = \frac{\lambda_+^n \lambda_-^n}{(\Delta c)^{2n}} \cdot \frac{(c_+ t - x)^n(x - c_- t)^{n-1}}{n!(n-1)!}\theta_t, \qquad (15)$$

$$n = 1, 2, \ldots$$

and

$$q_{2n+1}^{(\pm)} = \frac{\lambda_+^{n+1}\lambda_-^n}{(\Delta c)^{2n+1}} \cdot \frac{(c_+ t - x)^n(x - c_- t)^n}{(n!)^2}\theta_t, \qquad q_{2n+1}^{(\pm)} = \frac{\lambda_+^n \lambda_-^{n+1}}{(\Delta c)^{2n+1}} \cdot \frac{(c_+ t - x)^n(x - c_- t)^n}{(n!)^2}\theta_t, \qquad (16)$$

$$n = 0, 1, 2, \ldots$$

with $\theta_t = \theta_t(x) = \mathbf{1}_{\{c_- t < x < c_+ t\}}$.

This observation permits us to obtain the explicit distribution $p^{(\pm)}(x,t)\mathrm{d}x = \mathsf{P}_\pm\{X^{(\pm)}(t) \in \mathrm{d}x\}$:

$$p^{(\pm)}(x,t) = \sum_{n=0}^{\infty} p_n^{(\pm)}(x,t) = \mathrm{e}^{-\lambda_\pm t} \cdot \delta(x - c_\pm t)$$

$$+ \mathrm{e}^{-(\tilde{\lambda}t + \nu x)} \left[ \frac{\lambda_\pm}{\Delta c} I_0 \left( \frac{2\sqrt{\lambda_+ \lambda_-}}{\Delta c} \sqrt{(c_+ t - x)(x - c_- t)} \right) \right. \tag{17}$$

$$\left. + \frac{\sqrt{\lambda_+ \lambda_-}}{\Delta c} \cdot \left( \frac{c_+ t - x}{x - c_- t} \right)^{\pm 1/2} I_1 \left( \frac{2\sqrt{\lambda_+ \lambda_-}}{\Delta c} \sqrt{(c_+ t - x)(x - c_- t)} \right) \right].$$

Here $I_0(z) = \sum_{n=0}^{\infty} \frac{(z/2)^{2n}}{(n!)^2}$ and $I_1(z) = I_0'(z)$ are modified Bessel functions, $\tilde{\lambda} = \lambda_\pm - \nu c_\pm = \frac{\lambda_- c_+ - \lambda_+ c_-}{c_+ - c_-}$.

This formula coincides with the main result of [2] (see Theorem 4.1).

## 3.  Market model based on telegraph processes

Now we are ready to introduce the telegraph market model. The price of a risky asset $S(t)$ follows

$$dS(t) = S(t-)d\left(X^\sigma(t) + J^\sigma(t)\right), \ t > 0 \tag{18}$$

and the process $S(t)$, $t \geq 0$ is right-continuous. Here we let $X^\sigma = X^\sigma(t)$, $t \geq 0$ be a telegraph process with the states $(c_-, \lambda_-)$ and $(c_+, \lambda_+)$, $c_+ \geq c_-$, and $J^\sigma = J^\sigma(t) = \sum_{j=1}^{N^\sigma(t)} h_{\sigma(\tau_j-)}$ with $h_\pm > -1$. The initial state of the market is defined by $\sigma = \sigma(0)$.

The price of the non-risky asset (bank account) has the form

$$B(t) = \mathrm{e}^{Y^\sigma(t)}, \qquad Y^\sigma(t) = \int_0^t r_{\sigma(s)} ds, \tag{19}$$

where $r_-, r_+ > 0$.

According to the properties of stochastic exponentials, from (18) we have that

$$S(t) = S_0 \mathcal{E}_t \left( X^\sigma + J^\sigma \right) = S_0 \mathrm{e}^{X^\sigma(t)} \kappa^\sigma(t), \tag{20}$$

where $S_0 = S(0)$ and

$$\kappa^\sigma(t) = \prod_{s \leq t} (1 + \Delta J^\sigma(s)) = \kappa_{N^\sigma(t),\sigma}.$$

The sequence $\kappa_{n,\sigma}$, $n \geq 0$ is defined in (11) (with $h_\pm$ instead of $h_\pm^*$).

We assume that the parameters of the model (18)-(19) satisfy the conditions

$$\frac{r_\sigma - c_\sigma}{h_\sigma} > 0, \qquad \sigma = \pm 1. \tag{21}$$

Under such conditions, we can find a unique martingale measure in the framework of the market (18)-(19). Recall that the measure $\mathsf{P}^*_\sigma$, equivalent to $\mathsf{P}_\sigma$, is a martingale measure if the process $(B(t)^{-1}S(t))_{t\geq 0}$ is a $\mathsf{P}^*_\sigma$-martingale. We define this measure by the density $Z^\sigma(t)$, $t \geq 0$, (9), with $h^*_\pm = -c^*_\pm/\lambda_\pm$.

**Theorem 3.1:** *Measure $\mathsf{P}^*_\sigma$, defined by (9), is the martingale measure if and only if*

$$c^*_\sigma = \lambda_\sigma + \frac{c_\sigma - r_\sigma}{h_\sigma}, \quad \sigma = \pm 1.$$

*Moreover, under the probability measure $\mathsf{P}^*_\sigma$, the process $N^\sigma$ is a Poisson process with alternating intensities*

$$\lambda^*_\sigma = \frac{r_\sigma - c_\sigma}{h_\sigma} > 0, \qquad \sigma = \pm 1.$$

**Proof:** First of all, note that the condition (21) guarantees these inequalities: $h^*_\sigma = -c^*_\sigma/\lambda_\sigma = -1 + (r_\sigma - c_\sigma)/(\lambda_\sigma h_\sigma) > -1$ and $\lambda^*_\sigma = \lambda_\sigma - c^*_\sigma = (r_\sigma - c_\sigma)/h_\sigma > 0$, $\sigma = \pm 1$. Therefore, the process $Z^\sigma = Z^\sigma(t) = \mathcal{E}_t(X^\sigma_* + J^\sigma_*)$ with $c^*_\sigma = \lambda_\sigma + \frac{c_\sigma - r_\sigma}{h_\sigma}$ and $h^*_\sigma = -c^*_\sigma/\lambda_\sigma$ defines the density of the new probability measure correctly.

According to Theorem 2.4, the process $X^\sigma - Y^\sigma$ is a telegraph process (with respect to $\mathsf{P}^*_\sigma$) with the states $(c_\sigma - r_\sigma, \lambda_\sigma - c^*_\sigma)$, $\sigma = \pm 1$. From Theorem 2.3, it follows that $X^\sigma(t) - Y^\sigma(t) + J^\sigma(t)$, $t \geq 0$ is the $\mathsf{P}^*_\sigma$-martingale if and only if

$$(\lambda_\sigma - c^*_\sigma)h_\sigma = -(c_\sigma - r_\sigma).$$

Hence $c^*_\sigma = \lambda_\sigma + (c_\sigma - r_\sigma)/h_\sigma$ and $h^*_\sigma = -c^*_\sigma/\lambda_\sigma = -1 + (r_\sigma - c_\sigma)/\lambda_\sigma h_\sigma$, and the Theorem is proved. $\square$

To operate with assets $B$ and $S$, we will exploit the notion of a trading strategy (portfolio) as a pair of two predictable processes $\varphi_t$ and $\psi_t$, $t \geq 0$. Here $\varphi_t$ and $\psi_t$ are the amounts of the risky and the risk-free assets held in the portfolio at time $t$. The capital of the strategy is $F^{\varphi,\psi}_t = \varphi_t S(t) + \psi_t B(t)$. Such a strategy is self-financing if $dF^{\varphi,\psi}_t = \varphi_t dS(t) + \psi_t dB(t)$, and admissible if $F^{\varphi,\psi}_t \geq 0$, $t \geq 0$. Any self-financing strategy with a non-negative capital is called admissible. We will operate only with trading strategies which are admissible.

A trading strategy is called an arbitrage strategy (at time $T$) if its initial capital is zero and $\mathsf{P}_\sigma(F^{\varphi,\psi}_T > 0) > 0$. It is well-known (see Delbaen and Schachermayer [8]) that the existence of a martingale measure guarantees that the market does not admit arbitrage. Hence, according to Theorem 3.1, the market model (18)-(19) is arbitrage-free.

**Remark 1:** It is widely accepted that the telegraph process $X^\sigma$ has a persistent character. Therefore, if $J^\sigma(t) \equiv 0$, the market has arbitrage opportunities (see the model considered in Di Crescenzo and Pellerey [6]). The corresponding arbitrage strategy is described below.

Assume $r_\pm = 0$ for simplicity. Take numbers $A$, $B$ such that $S_0 < A < B < S_0 e^{c_+ T}$. Consider the following strategy: buy the risky asset at time $\mathfrak{t}_1 = \min\{t \in [0, T] : S(t) = A\}$, and then sell it at time $\mathfrak{t}_2 = \min\{t \in (\mathfrak{t}_1, T] : S(t) = A \text{ or } S(t) = B\}$. This strategy has no losses at time $\mathfrak{t}_1$, because $\mathfrak{t}_1$ coincides with the switching time of $X$ with zero probability. Hence the strategy creates a positive profit with positive probability $\mathsf{P}\{S(\mathfrak{t}_2) = B\}$.

Now, let us discuss the convergence of (18) to the Black-Scholes model. First M.Kac [14] noticed that the telegraph equation tends to the heat equation when $c$, $\lambda \to \infty$ and $c^2/\lambda \to 1$ (see also E.Orsingher [19], [20]). Moreover the homogeneous telegraph process $X(\cdot)$ converges in distribution, as $c$, $\lambda \to \infty$, and $c^2/\lambda \to 1$, to the standard Brownian motion $w(\cdot)$ in $C([0, T]; (-\infty, \infty))$ (equipped with the sup-norm and the $\sigma$-algebra generated by the open subsets). See details in N.Ratanov [21].

The following theorem provides a similar connection between stock prices driven by geometric telegraph processes and geometric Brownian motion. Seeking for simplicity we consider the symmetric case $\lambda_- = \lambda_+$, $c_- = a - c$, $c_+ = a + c$.

**Theorem 3.2:** *Let $\lambda_- = \lambda_+ = \lambda \to \infty$, $c \to \infty$,*

$$c^2/\lambda \to v_c^2 \qquad a^2/\lambda \to v_a^2. \tag{22}$$

*Let $h_-$, $h_+ \to 0$ and*

$$a + \lambda B/2 \to \mu, \tag{23}$$

*where $B = \ln\left[(1 + h_-)(1 + h_+)\right]$.*

*Then*

$$S(\cdot) \xrightarrow{D} S_0 \exp\left(vw(\cdot) + \mu \cdot\right), \tag{24}$$

*where $\xrightarrow{D}$ means convergence in distribution in $C([0, T]; (-\infty, \infty))$, and $v = \sqrt{v_c^2 + v_a^2}$.*

**Proof:** Let $f_\pm(z, t) = \mathsf{E}_\pm \mathrm{e}^{z(X^{(\pm)}(t) + \ln \kappa^{(\pm)}(t))}$ be the moment-generating function of jump telegraph process $X^{(\pm)}(t) + \ln \kappa^{(\pm)}(t)$, where $\kappa^{(\pm)}(t) = \kappa_{n,\pm}$ if $N^{(\pm)}(t) = n$ (sequence $\kappa_{n,\pm}$, $n \geq 0$ is defined by (11) with $h_\pm$ instead of $h_\pm^*$). We prove here the convergence

$$f_\pm(z, t) \to \exp(\mu z t + v^2 z^2 t/2), \tag{25}$$

which is sufficient for the convergence of one-dimensional distributions in (24). From Lemma 2.2 it follows that

$$f_\pm(z, t) = \mathsf{E}_\pm \mathrm{e}^{z\left(cX_{\mathrm{st}}^{(\pm)}(t) + at + \ln \kappa^{(\pm)}(t)\right)} = \mathrm{e}^{azt} \sum_{n=0}^{\infty} \int_{-\infty}^{\infty} \mathrm{e}^{z(xc + \ln \kappa_{n,\pm})} p_n^{(\pm)}(x, t) \mathrm{d}x, \tag{26}$$

where $X_{\mathrm{st}}^{(\pm)}$ denotes a standard telegraph process with states $(\pm 1, \lambda)$, and $p_n^{(\pm)}$, $n \geq 0$ here denote the (generalized) probability densities of $X_{\mathrm{st}}^{(\pm)}(t)$: $p_n^{(\pm)}(x, t) = \mathrm{e}^{-\lambda t} q_n^{(\pm)}(x, t)$, where $q_n^{(\pm)}$ are defined in (15)-(16) with $c_\pm = \pm 1$ and $\lambda_\pm = \lambda$.

As $h_\pm \to 0$

$$f_\pm(z, t) \sim \mathrm{e}^{azt} \int_{-\infty}^{\infty} \mathrm{e}^{czx} \sum_{n=0}^{\infty} \mathrm{e}^{znB/2} p_n^{(\pm)}(x, t) \mathrm{d}x \tag{27}$$

Here and below $f \sim g$ denotes the existence and equality of limits of $f$ and $g$ under scaling (22)-(23).

Taking into account formulae (15)-(16) and changing variables in the integral in (26), $x' = cx$, we obtain

$$f_\pm(z, t) \sim \mathrm{e}^{azt + (\bar{\lambda} - \lambda)t} \int_{-\infty}^{\infty} \mathrm{e}^{zx} \bar{p}^{(\pm)}(x, t) dx,$$

where $\bar{p}^{(\pm)}(x, t)$ is the density of the telegraph process $\bar{X}^{(\pm)}(t)$ with states $(\pm c, \bar{\lambda})$, $\bar{\lambda} = \lambda \mathrm{e}^{zB/2}$.

Next, note that

$$\bar{\lambda} - \lambda + az = \lambda(e^{zB/2} - 1) + az$$

$$\sim \frac{\lambda zB}{2} + \frac{\lambda z^2 B^2}{8} + az.$$

From (22)-(23), it follows that $\sqrt{\lambda}B/2 \sim -a/\sqrt{\lambda}$ and

$$\bar{\lambda} - \lambda + az \to \mu z + v_a^2 z^2/2.$$

It is well-known [14] that under the scaling $c^2/\lambda \to v^2$ the process $\bar{X}(t)$ converges in distribution to diffusion process $v \cdot w(t)$, and thus densities $\bar{p}^{(\pm)}(\cdot, t)$ converge to the probability density of $v_c w(t)$ (see e.g. [3]):

$$\bar{p}^{(\pm)}(x, t) \to \frac{1}{v_c \sqrt{2\pi t}} e^{-x^2/2tv_c^2}.$$

Summarizing the above statements, we obtain (25). The convergence of any finite-dimensional distributions and the compactness property could be proved in the same manner as in N.Ratanov [21]. $\square$

**Remark 2:** Condition (23) in this theorem means that the total drift $a + \lambda B/2$ is asymptotically finite. Here $a = (c_- + c_+)/2$ is generated by the velocities of telegraph process $X$, and the summand $\lambda B/2$ represents the drift component (possibly with infinite asymptotics) that is provoked by jumps. If in (23) the limit of $\lambda B/2$ is finite, then $a \to \alpha \equiv const$, and in (24) the drift volatility term $v_a = 0$.

In general, by (22)-(23), we have that $\sqrt{\lambda}B/2 \to -v_a$, so $-\sqrt{\lambda}B/2$ represents the jump component of volatility.

### 4. Fundamental equation and perfect hedging

Consider a European option with maturity time $T$ and payoff function $f(S(T))$. We assume $f$ is a continuous and piecewise smooth function. To price these options, we need to study the function

$$F(t, x, \sigma) = \mathsf{E}_\sigma^* \left[ e^{-Y^\sigma(T-t)} f(x e^{X^\sigma(T-t)} \kappa^\sigma(T-t)) \right], \qquad (28)$$

$$\sigma = \pm 1, \ 0 \le t \le T,$$

where $\mathsf{E}_\sigma^*$ denotes the expectation with respect to the martingale measure $\mathsf{P}_\sigma^*$, under the initial state $\sigma = \sigma(0)$ of the underlying Markov process $\sigma(t)$, $t \ge 0$, $\sigma = \pm 1$.

**Theorem 4.1:** *Function $F$ is a solution of the following hyperbolic system*

$$\frac{\partial F}{\partial t}(t,x,\sigma) + c_\sigma x \frac{\partial F}{\partial x}(t,x,\sigma)$$

$$= (r_\sigma + \frac{r_\sigma - c_\sigma}{h_\sigma})F(t,x,\sigma) - \frac{r_\sigma - c_\sigma}{h_\sigma}F(t, x(1+h_\sigma), -\sigma), \ 0 < t < T, \quad \sigma = \pm 1 \qquad (29)$$

*with the terminal condition $F(T, x, \sigma) = f(x)$.*

**Proof:** Note that $Y^\sigma(t) = a_r X^\sigma(t) + b_r t$ with $a_r = \frac{r_+ - r_-}{c_+ - c_-}$, $b_r = \frac{c_+ r_- - c_- r_+}{c_+ - c_-}$ (see Lemma 2.2). Conditioning on the number of jumps, we can write

$$F(t,\ x,\ \sigma) = e^{-b_r(T-t)} \sum_{n=0}^{\infty} \int_{-\infty}^{\infty} e^{-a_r y} f(x e^y \kappa_{n,\sigma}) p_{*,n}^{(\sigma)}(y,\ T-t) dy, \quad \sigma = \pm 1, \tag{30}$$

where $p_{*,n}^{(\pm)}$, $n \geq 0$ are the (generalized) probability densities of the telegraph process $X^\pm = X^\pm(t)$, $0 \leq t \leq T$ (with $n$ turns), with respect to the martingale measure $\mathsf{P}_\pm^*$. The densities $p_{*,n}^{(\pm)}$ have a form (13)-(16) (with $\lambda_\pm^*$ instead of $\lambda_\pm$ and the same coefficient $c_\pm$). Therefore the series in (30) (and its derivatives) uniformly converges. It permits us to apply the equations (14). Taking into account the identity $c_\sigma a_r + b_r = r_\sigma$, $\sigma = \pm 1$ (see Lemma 2.2), we obtain from (30)

$$\frac{\partial F}{\partial t}(t,\ x,\ \sigma) + c_\sigma x \frac{\partial F}{\partial x}(t,\ x,\ \sigma)$$

$$= (r_\sigma + \lambda_\sigma^*) F(t,\ x,\ \sigma) - \lambda_\sigma^* e^{-b_r(T-t)} \sum_{n=1}^{\infty} \int_{-\infty}^{\infty} e^{-a_r y} f(x e^y \kappa_{n,\sigma}) p_{*,n-1}^{(-\sigma)}(y,\ T-t) dy.$$

By the equalities (11) and $\lambda_\sigma^* = \frac{r_\sigma - c_\sigma}{h_\sigma}$, the latter equation becomes (29). □

**Remark 1:** The system (29) plays the same role for our model as the fundamental Black-Scholes equation. In contrast with classical theory, this system is hyperbolic. In particular, it implies the finite velocity of propagation, which corresponds better to the intuitive understanding of financial markets. Note that the equations (29) do not depend on $\lambda_\pm$, just as the respective equation in the Black-Scholes model does not depend on the drift parameter.

Now we consider the hedging problem for the option with a payoff function $\mathcal{H}$, which is $\mathbf{F}_T$-measurable. The self-financing strategy $\pi = (\varphi_t,\ \psi_t)$, $0 \leq t \leq T$ is called a hedge (perfect hedge, replicating strategy) if its terminal value is equal to the payoff of the option:

$$F_T^\pi = \mathcal{H} \qquad \mathsf{P}-\text{a.s.} \tag{31}$$

For the wealth process $F_t = F_t^\pi$, we require that

$$F_t = \varphi_t S(t) + \psi_t B(t),\ 0 \leq t \leq T \tag{32}$$

and

$$dF_t = \varphi_t dS(t) + \psi_t dB(t). \tag{33}$$

Let us rewrite (33) in the integral form

$$F_t = F_0 + \int_0^t \varphi_s S(s) dX^\sigma(s) + \int_0^t \psi_s dB(s) + \sum_{j=1}^{N^\sigma(t)} \varphi_{\tau_j} h_{\sigma(\tau_j-)} S(\tau_j-).$$

Using the equality $\psi_t = B(t)^{-1}(F_t - \varphi_t S(t))$ (see the balance equation (32)), we can rewrite the above

equation as

$$F_t = F_0 + \int_0^t r_{\sigma(s)} F_s ds$$

$$+ \int_0^t \varphi_s S(s)(c_{\sigma(s)} - r_{\sigma(s)}) ds + \sum_{j=1}^{N^\sigma(t)} \varphi_{\tau_j} h_{\sigma(\tau_j-)} S(\tau_j-). \tag{34}$$

To identify such a strategy in the case $\mathcal{H} = f(S(T))$, note that $F_t = F_t^\pi = B(t) \mathsf{E}_\sigma^*[B(T)^{-1}\mathcal{H} \mid \mathbf{F}_t] = F(t, S(t), \sigma(t))$, where $F(t, x, \sigma)$ is defined by (28) and satisfies the fundamental equation (29). Exploiting Ito's formula, generalized by Dolean and Meyer (see e. g. [17]), we get

$$F_t = F_0 + \int_0^t \frac{\partial F}{\partial s}(s, S(s), \sigma(s)) ds + \int_0^t \frac{\partial F}{\partial x}(s, S(s), \sigma(s)) S(s) c_{\sigma(s)} ds \tag{35}$$

$$+ \sum_{j=1}^{N^\sigma(t)} (F_{\tau_j} - F_{\tau_j-}).$$

Comparing the latter two equations and utilizing the fundamental equation (29), we have (between jumps)

$$\varphi_t = \frac{S(t) c_{\sigma(t)} \frac{\partial F}{\partial x} + \frac{\partial F}{\partial t} - r_{\sigma(s)} F}{S(t)(c_{\sigma(t)} - r_{\sigma(s)})}$$

$$= \frac{F(t, S(t)(1 + h_{\sigma(t)}), -\sigma(t)) - F(t, S(t), \sigma(t))}{S(t) h_{\sigma(t)}}. \tag{36}$$

Moreover, from (34) and (35), we obtain the values of $\varphi_{\tau_j}$:

$$\varphi_{\tau_j} = \frac{F_{\tau_j} - F_{\tau_j-}}{S(\tau_j-) h_{\sigma(\tau_j-)}} = \frac{F(\tau_j, S(\tau_j), \sigma(\tau_j)) - F(\tau_j, S(\tau_j-), -\sigma(\tau_j))}{S(\tau_j-) h_{\sigma(\tau_j-)}}. \tag{37}$$

It turns out that the process $\varphi_t$ is left-continuous. To prove this, we note that based on (20),

$$S(\tau_j-)(1 + h_{\sigma(\tau_j-)}) = S(\tau_j). \tag{38}$$

Now it is sufficient to apply (38) to (36)-(37).

## 5. Pricing call options

The main goal of this section is to derive an exact formula for the initial price $\mathfrak{c}$ of a call option with payoff $(S(T) - K)^+$ in the framework of the market (18)-(19). According to the theory on option pricing (see for example Karatzas and Shreve [16], Duffie [9]), we have

$$\mathfrak{c}^\sigma = \mathsf{E}_\sigma^*(B(T)^{-1}(S(T) - K)^+),$$

where $K$ is the strike price and $\mathsf{E}_\sigma^*(\cdot)$ is the expectation with respect to the martingale measure $\mathsf{P}_\sigma^*$.

In case of the markovian model (19)-(20), one can rewrite $\mathfrak{c}^\sigma$ as

$$\mathfrak{c} = \mathfrak{c}^\sigma = \sum_{n=0}^{\infty} \mathsf{E}_\sigma^* (B(T)^{-1}(S(T) - K)^+ \mathbf{1}_{\{N^\sigma(T)=n\}}), \tag{39}$$

where $\sigma = \pm 1$ indicates the initial state.

We rewrite (39) in the form

$$\mathfrak{c}^\sigma = S_0 U^{(\sigma)}(y, T) - K u^{(\sigma)}(y, T) \tag{40}$$

with

$$u^{(\sigma)}(y, T) = \sum_{n=0}^{\infty} u_n^{(\sigma)}(y - b_n^{(\sigma)}, T), \qquad U^{(\sigma)}(y, T) = \sum_{n=0}^{\infty} U_n^{(\sigma)}(y - b_n^{(\sigma)}, T),$$

where $y = \ln K/S_0$, $b_n^{(\sigma)} = \ln \kappa_{n,\sigma}$, and functions $u_n^{(\sigma)}$, $U_n^{(\sigma)}$, $n \geq 0$ are defined as follows:

$$u_n^{(\sigma)}(y,\, t) = u_n^{(\sigma)}(y,\, t;\, \lambda_\pm^*,\, c_\pm,\, r_\pm) = \mathsf{E}_\sigma^* \left[B(t)^{-1}\mathbf{1}_{\{X^\sigma(t)>y,\, N^\sigma(t)=n\}}\right] \tag{41}$$

$$= e^{-b_r t} \int_y^\infty e^{-a_r x} p_{*,n}^{(\sigma)}(x,\, t) dx$$

with $a_r = \frac{r_+ - r_-}{c_+ - c_-}$ and $b_r = \frac{c_+ r_- - c_- r_+}{c_+ - c_-}$;

$$U_n^{(\sigma)}(y,\, t) = U_n^{(\sigma)}(y,\, t;\, \lambda_\pm^*,\, c_\pm,\, r_\pm) = \mathsf{E}_\sigma^* \left[B(t)^{-1}\mathcal{E}_t(X^\sigma + J^\sigma)\mathbf{1}_{\{X^\sigma(t)>y,\, N^\sigma(t)=n\}}\right] \tag{42}$$

$$= \kappa_{n,\sigma} e^{-b_r t} \int_y^\infty e^{-a_r x + x} p_{*,n}^{(\sigma)}(x,\, t) dx.$$

The functions $u_n^{(\sigma)}(y,\, t)$, $n \geq 1$ satisfy the equations (see (14))

$$\frac{\partial u_n^{(\sigma)}}{\partial t}(y,\, t) + c_\sigma \frac{\partial u_n^{(\sigma)}}{\partial y}(y,\, t) = -(\lambda_\sigma^* + r_\sigma) u_n^{(\sigma)}(y,\, t) + \lambda_\sigma^* u_{n-1}^{(-\sigma)}(y,\, t) \tag{43}$$

with initial conditions $u_n^{(\sigma)}|_{t=0} = 0$, $n \geq 1$. As it is easy to see exploiting Remark **2** of Section 2, these functions ($u_n^{(\sigma)}$, $n \geq 1$) are continuous and piece-wise continuously differentiable, and $u_0^{(\sigma)}(y,t) = e^{-(\lambda_\sigma^* + r_\sigma)t}\theta(c_\sigma t - y)$, $\sigma = \pm 1$. Moreover, $\forall n, u_n^{(\sigma)} \equiv 0$ if $y > c_+ t$, and

$$u_n^{(\sigma)}(y,t) \equiv \rho_n^{(\sigma)}(t) = e^{-b_r t} \int_{-\infty}^\infty e^{-a_r x} p_{n,*}^{(\sigma)}(x,\, t) dx \tag{44}$$

if $y < c_- t$. In the latter case, the system (43) takes the form

$$\frac{d\rho_n^{(\sigma)}}{dt} = -(\lambda_\sigma^* + r_\sigma)\rho_n^{(\sigma)} + \lambda_\sigma^* \rho_{n-1}^{(-\sigma)}, \quad n \geq 1, \tag{45}$$

$\rho_0^{(\sigma)} = e^{-(\lambda_\sigma^* + r_\sigma)t}$ and $\rho_n^{(\sigma)}|_{t=0} = 0$, $n \geq 1$, $\sigma = \pm 1$.

**Lemma 5.1:** *The solution of system (45) can be represented in the form*

$$\rho_n^{(\sigma)}(t) = e^{-(\lambda_-^* + r_-)t} \Lambda_n^{(\sigma)} P_n^{(\sigma)}(t), \quad \sigma = \pm 1, \ n \geq 0,$$

*where* $\Lambda_n^{(\sigma)} = (\lambda_\sigma^*)^{[(n+1)/2]} (\lambda_{-\sigma}^*)^{[n/2]}$ *and functions* $P_n^{(\sigma)}$ *are defined as follows:*

$$P_0^{(+)} = e^{-at}, \qquad P_0^{(-)} \equiv 1,$$

$$P_n^{(\sigma)} = P_n^{(\sigma)}(t) = \frac{t^n}{n!} \left[ 1 + \sum_{k=1}^{\infty} \frac{(m_n^{(\sigma)} + 1)_k}{(n+1)_k} \cdot \frac{(-at)^k}{k!} \right], \quad \sigma = \pm 1, \ n \geq 1. \quad (46)$$

*Here*

$$m_n^{(+)} = [n/2], \ m_n^{(-)} = [(n-1)/2],$$

$$(m)_k = m(m+1) \ldots (m+k-1), \quad a = \lambda_+^* - \lambda_-^* + r_+ - r_-.$$

**Proof:** Notice that in the particular case $\lambda_+^* = \lambda_-^* = \lambda$ and $r_\pm = 0$, the solution of system (45) is well known: $\rho_n^{(\pm)}(t) = \pi_n(t) = \mathsf{P}(N(t) = n) = \frac{(\lambda t)^n}{n!} e^{-\lambda t}$.

Generally, we apply the following change of variables

$$\rho_n^{(\sigma)}(t) = e^{-(\lambda_-^* + r_-)t} \Lambda_n^{(\sigma)} P_n^{(\sigma)}(t).$$

In these notations, we have $P_0^{(+)}(t) = e^{-at}$, $a = (\lambda_+^* + r_+) - (\lambda_-^* + r_-)$; $P_0^{(-)}(t) = 1$; $P_n^{(\pm)}|_{t=0} = 0$, $n \geq 1$ and the system

$$\begin{cases} \dot{P}_n^{(+)} + a P_n^{(+)} = P_{n-1}^{(-)} \\ \dot{P}_n^{(-)} = P_{n-1}^{(+)} \end{cases}, \quad n \geq 1, \quad (47)$$

$\dot{P}_n^{(\pm)} = \frac{dP_n^{(\pm)}}{dt}$.

The latter system has the following solution

$$P_{2n+1} \equiv P_{2n+1}^{(\pm)} = \frac{t^{2n+1}}{(2n+1)!} \left[ 1 + \sum_{k=1}^{\infty} \frac{(n+1)\ldots(n+k)}{(2n+2)\ldots(2n+k+1)} \cdot \frac{(-at)^k}{k!} \right],$$

$$P_{2n}^{(-)} = \frac{t^{2n}}{(2n)!} \left[ 1 + \sum_{k=1}^{\infty} \frac{n(n+1)\ldots(n+k-1)}{(2n+1)\ldots(2n+k)} \cdot \frac{(-at)^k}{k!} \right],$$

$$P_{2n}^{(+)} = \frac{t^{2n}}{(2n)!} \left[ 1 + \sum_{k=1}^{\infty} \frac{(n+1)\ldots(n+k)}{(2n+1)\ldots(2n+k)} \cdot \frac{(-at)^k}{k!} \right],$$

which coincides with (46). □

**Remark 1:** Formulas (46) can be expressed using hypergeometric functions (Abramowitz and Stegun [1]):

$$P_n^{(\sigma)}(t) = \frac{t^n}{n!} \cdot {}_1F_1(m_n^{(\sigma)} + 1; \ n + 1; \ -at), \ m_n^{(+)} = [n/2], \ m_n^{(-)} = [(n-1)/2].$$

A hypergeometric function $_1F_1(\alpha;\ \beta;\ z)$ is defined as

$$_1F_1(\alpha;\ \beta;\ z) = 1 + \sum_{n=1}^{\infty} \frac{\alpha(\alpha+1)\ldots(\alpha+n-1)}{n!\beta(\beta+1)\ldots(\beta+n-1)} z^n = 1 + \sum_{n=1}^{\infty} \frac{(\alpha)_n}{n!(\beta)_n} z^n.$$

Also using (48), one can easily check that $P_{2n}^{(-)} - P_{2n}^{(+)} = aP_{2n+1},\ n \geq 0$.

Let us define the coefficients $\beta_{k,j},\ j < k$: $\beta_{k,0} = \beta_{k,1} = \beta_{k,k-2} = \beta_{k,k-1} = 1$,

$$\beta_{k,j} = \frac{(k-j)_{[j/2]}}{[j/2]!}, \tag{48}$$

and the functions $\varphi_{k,n}$: $\varphi_{0,n} = P_{2n+1}$ and

$$\varphi_{k,n} = \sum_{j=0}^{k-1} a^{k-j-1} \beta_{k,j} P_{2n-j}^{(-)},\ 1 \leq k \leq n. \tag{49}$$

For positive $p,\ q$, we define $v_0^{(-)} \equiv 0$, $v_0^{(+)} = e^{-ap}$, $v_1^{(\sigma)} = P_1(p)$, $\sigma = \pm 1$, and for $n \geq 1$

$$\begin{aligned} v_{2n+1}^{(\pm)} &= v_{2n+1}^{(\pm)}(p,\ q) = P_{2n+1}(p) + \sum_{k=1}^{n} \frac{q^k}{k!} \varphi_{k,n}(p), \\ v_{2n}^{(-)} &= v_{2n}^{(-)}(p,\ q) = P_{2n}^{(-)}(p) + \sum_{k=1}^{n-1} \frac{q^k}{k!} \varphi_{k+1,n}(p), \\ v_{2n}^{(+)} &= v_{2n}^{(+)}(p,\ q) = P_{2n}^{(+)}(p) + \sum_{k=1}^{n} \frac{q^k}{k!} \varphi_{k-1,n-1}(p). \end{aligned} \tag{50}$$

Now we can find expressions for $u_n^{(\sigma)} = u_n^{(\sigma)}(y,\ t)$ in the interval $c_- t < y < c_+ t$.

**Theorem 5.2:** *System (43) admits a unique solution of the form*

$$u_n^{(\sigma)} = \begin{cases} 0,\ y > c_+ t, \\ w_n^{(\sigma)}(p,\ q),\ c_- t \leq y \leq c_+ t, \quad \sigma = \pm 1, \\ \rho_n^{(\sigma)}(t),\ y < c_- t, \end{cases} \tag{51}$$

*where* $w_n^{(\sigma)} = e^{-(\lambda_+^* + r_+)q - (\lambda_-^* + r_-)p} \Lambda_n^{(\sigma)} v_n^{(\sigma)}(p,\ q)$, $p = \frac{c_+ t - y}{c_+ - c_-}$, $q = \frac{y - c_- t}{c_+ - c_-}$, $n \geq 0$; *functions $\rho_n^{(\sigma)}$ are expressed in Lemma 5.1.*

**Proof:** Evidently, $u_n^{(\sigma)}(y,\ t) \equiv 0$, if $p < 0$, and $u_n^{(\sigma)}(y,\ t) \equiv \rho_n^{(\sigma)}(t)$, if $q < 0$. For $p,\ q > 0$ we have the system

$$\begin{cases} \frac{\partial v_n^{(+)}}{\partial q} = v_{n-1}^{(-)}, \\ \frac{\partial v_n^{(-)}}{\partial p} = v_{n-1}^{(+)} \end{cases},\quad n \geq 1 \tag{52}$$

with

$$v_0^{(+)} = e^{-ap}\theta(p),\quad v_0^{(-)} = e^{aq}\theta(-q),\quad v_n^{(\pm)}|_{p<0} \equiv 0$$

and

$$v_n^{(\sigma)} \mid_{q<0} = e^{aq} P_n^{(\sigma)}(p+q). \tag{53}$$

Here $a = (\lambda_+^* + r_+) - (\lambda_-^* + r_-)$ and $P_n^{(\sigma)}$, $n \geq 0$, $\sigma = \pm 1$ are defined in (46).

It is straightforward to check that the exact representation of the solution of (52) for $p$, $q > 0$ has the form (50) with $\varphi_{0,n} = P_{2n+1}$, $\varphi_{1,n} = P_{2n}^{(-)}$ and

$$\varphi'_{k,n} = \varphi_{k-1,n-1}, \ 1 \leq k \leq n. \tag{54}$$

The proof is finished by the following proposition.

**Proposition 5.3:** *The solution of the system (54) has the form (49):*
$$\varphi_{k,n} = \sum_{j=0}^{k-1} a^{k-j-1} \beta_{k,j} P_{2n-j}^-.$$

*Proof.* Indeed, from (49) and (47) it follows that

$$\varphi'_{k,n} = \sum_{j=0}^{k-1} a^{k-j-1} \beta_{k,j} P_{2n-j-1}^{(+)}.$$

By the identities $P_{2n+1}^{(+)} = P_{2n+1}^{(-)}$ and $P_{2n}^{(-)} - P_{2n}^{(+)} = aP_{2n+1}$, $n \geq 0$ (see Remark **1** of this Section), we have

$$\varphi'_{k,n} = \sum_{j \geq 0, \ j \text{ is even}} a^{k-j-1} \beta_{k,j} P_{2n-j-1}$$

$$+ \sum_{j \geq 0, \ j \text{ is odd}} a^{k-j-1} \beta_{k,j} P_{2n-j-1}^{(-)} - \sum_{j \geq 0, \ j \text{ is odd}} a^{k-j} \beta_{k,j} P_{2n-j}.$$

To complete the proof, it is sufficient to apply the identities $\beta_{k,2m+1} = \beta_{k-1,2m}$, $\beta_{k,2m} - \beta_{k,2m+1} = \beta_{k-1,2m-1}$, which are evident from the definition of $\beta_{k,n}$ (see (48)). $\square$

**Remark 2:** If $\lambda_-^* = \lambda_+^* = \lambda$, $r_+ = r_- = r$, then $P_n^{(\sigma)} = \frac{t^n}{n!}$, $\pi_n^{(\sigma)} \equiv \pi_n = \frac{(\lambda t)^n}{n!} e^{-\lambda t}$, $\rho_n^{(\sigma)} = e^{-rt} \pi_n(t)$ and $\varphi_{k,n} = P_{2n-k+1}^{(\sigma)} = \frac{t^{2n-k+1}}{(2n-k+1)!}$. Moreover,

$$v_n^{(\sigma)} = \frac{1}{n!} \sum_{k=0}^{m_n^{(\sigma)}} \binom{n}{k} q^k p^{n-k}.$$

**Remark 3:** It follows from (51) that functions $u_0^{(-)}$ and $u_0^{(+)}$ are discontinuous at $q = 0$ and $p = 0$ respectively. All other functions $u_n^{(\sigma)}$, $n \geq 1$, defined in (51), are continuous. The points of possible discontinuity of derivatives are concentrated on the lines $p = 0$ and $q = 0$. For example, for $u_1^{(\sigma)}$ and $\sigma = \pm 1$, we have

$$\frac{\partial u_1^{(\sigma)}}{\partial q} \mid_{q=+0} - \frac{\partial u_1^{(\sigma)}}{\partial q} \mid_{q=-0} = \lambda_\sigma^* e^{-(\lambda_+^* + r_+)p}$$

and

$$\frac{\partial u_1^{(\sigma)}}{\partial p}\Big|_{p=+0} - \frac{\partial u_1^{(\sigma)}}{\partial p}\Big|_{p=-0} = \lambda_\sigma^* e^{-(\lambda_+^* + r_+)q}.$$

Moreover, using (51) one can prove that $u_n^{(\sigma)} \in \mathcal{C}^{n-1}$.

Similarly to system (43) for $u_n^{(\sigma)}$, the functions $U_n^{(\sigma)} = U_n^{(\sigma)}(y,\ t)$, $n \geq 1$, defined in (42), satisfy the equations

$$\frac{\partial U_n^{(\sigma)}}{\partial t} + c_\sigma \frac{\partial U_n^{(\sigma)}}{\partial y} = -(\lambda_\sigma^* + r_\sigma - c_\sigma)U_n^{(\sigma)} + \lambda_\sigma^*(1+h_\sigma)U_{n-1}^{(-\sigma)}. \tag{55}$$

Hence we obtain the following representation:

$$U_n^{(\sigma)}(y,\ t;\ \lambda_\pm^*,\ c_\pm,\ r_\pm) = u_n^{(\sigma)}(y,\ t;\ \bar\lambda_\pm,\ c_\pm,\ 0), \tag{56}$$

where $\bar\lambda_\sigma = \lambda_\sigma^*(1+h_\sigma) = \lambda_\sigma^* + r_\sigma - c_\sigma$.

**Remark 4:** The formulas in (40) have a different structure, which depends on the sign of $\ln(1+h_-)(1+h_+)$.

(i) If $(1+h_-)(1+h_+) < 1$, then $\ln(1+h_-) + \ln(1+h_+) < 0$ and $b_n^{(\sigma)} \to -\infty$. The price of a call option is given by the formula (40) with

$$u = u^{(\sigma)}(y,\ T) = \sum_{k=0}^{n_-^{(\sigma)}} \rho_k^{(\sigma)}(T) + \sum_{k=n_-^{(\sigma)}+1}^{n_+^{(\sigma)}} u_k^{(\sigma)}(y - b_k^{(\sigma)},\ T;\ \lambda_\pm^*,\ c_\pm,\ r_\pm),$$

and

$$U = U^{(\sigma)}(y,\ T) = u^{(\sigma)}(y,\ T;\ \bar\lambda_\pm,\ c_\pm,\ 0), \tag{57}$$

where $y = \ln K/S_0$ and $n_-^{(\sigma)} = \min\left\{n\ :\ y - b_n^{(\sigma)} > c_- T\right\}$, $n_+^{(\sigma)} = \min\left\{n\ :\ y - b_n^{(\sigma)} > c_+ T\right\}$.

(ii) If $(1+h_-)(1+h_+) > 1$, then $\ln(1+h_-) + \ln(1+h_+) > 0$ and $b_n^{(\sigma)} \to +\infty$. Denoting

$$m_-^{(\sigma)} = \max\left\{n\ :\ y - b_n^{(\sigma)} > c_- T\right\},$$

$$m_+^{(\sigma)} = \max\left\{n\ :\ y - b_n^{(\sigma)} > c_+ T\right\},$$

we obtain the call option price formula of the form (40) with

$$u^{(\sigma)}(y,\ T) = \sum_{k=m_+^{(\sigma)}}^{m_-^{(\sigma)}} u_k^{(\sigma)}(y - b_k^{(\sigma)},\ T;\ \lambda_\pm^*,\ c_\pm,\ r_\pm) + \sum_{k=m_-^{(\sigma)}+1}^{\infty} \rho_k^{(\sigma)}(T),$$

and $U^{(\sigma)}(y,\ T)$ is defined in (57).

Consider the following examples.

**Example 5.4** The Merton model.

If $r_- = r_+ = r$, $c_- = c_+ = c$, $h_- = h_+ = -h$, $\lambda_- = \lambda_+ = \lambda$, equation (18) has the form

$$dS(t) = S(t-)(cdt - hdN(t)),$$

where $N = N(t)$, $t \geq 0$ is a (homogeneous) Poisson process with parameter $\lambda > 0$. In this case, the formula (40) can be simplified to

$$\mathfrak{c} = S_0 U(\ln K/S_0, \ T) - K u(\ln K/S_0, \ T). \tag{58}$$

Here functions $u$ and $U$ are defined as follows.

If $0 < h < 1$ and $c > r$, then $b_n^{(\sigma)} \equiv b_n = n \ln(1-h) \downarrow -\infty$ and

$$u = u(\ln K/S_0, \ T) = e^{-rT} \sum_{n=0}^{n_0} u_n^{(\sigma)}(\ln(K/S_0) - b_n, \ T)$$

$$= e^{-rT} \mathsf{P}(N(T) \leq n_0) = e^{-rT} \Psi_{n_0}(\lambda^* T),$$

where $\lambda^* = (c-r)/h > 0$ and $\Psi_{n_0}(z) = e^{-z} \sum_{n=0}^{n_0} \frac{z^n}{n!}$. In this case, the function $U$ has the form

$$U(y, \ T) = \Psi_{n_0}(\lambda^*(1-h)T).$$

For $h < 0$ and $c < r$, i. e. $b_n^{(\sigma)} = n \ln(1-h) \uparrow +\infty$, we have

$$u(y, \ T) = e^{-rT}\left(1 - \Psi_{n_0}(\lambda^* T)\right),$$

$$U(y, \ T) = 1 - \Psi_{n_0}(\lambda^*(1-h)T).$$

In both cases,

$$n_0 = \inf\{n : S_0 e^{n \ln(1-h)+(c-r)T} > B(T)^{-1}K\} = \left[\frac{\ln(K/S_0) - cT}{\ln(1-h)}\right].$$

**Example 5.5** Let us consider another symmetric case $\lambda_+ = \lambda_- := \lambda$, $r_+ = r_- := r$, $c_+ = r+c, c_- = r-c$ and $h_+ = -h, h_- = h$; $c > 0$, $0 < h < 1$. These assumptions simplify the form of $u^{(\sigma)}$. In this case we have $\lambda_+^* = \lambda_-^* = c/h$ and $b_n^{(\pm)} \to -\infty$. Here $b_{2n}^{(\pm)} = n \ln(1-h^2)$ and $b_{2n+1}^{(\pm)} = n \ln(1-h^2) + \ln(1 \mp h)$. We denote

$$n_\pm = \left[\frac{\ln(K/S_0) - (c \pm r)T}{\ln(1-h^2)}\right]. \tag{59}$$

Function $u^{(+)}$ has the form (see Remark **2** of this Section)

$$u^{(+)}(y, T) = e^{-(c/h+r)T}\left[\sum_{n=0}^{2n_-} \frac{(cT/h)^n}{n!} + \sum_{n=2n_-+1}^{2n_+} \frac{(c/h)^n}{n!} \sum_{k=0}^{[n/2]} \binom{n}{k} q_n^k p_n^{n-k}\right],$$

where $p_n = \frac{(r+c)T-y+b_n^{(+)}}{2c} \geq 0$, $q_n = \frac{y-(r-c)T-b_n^{(+)}}{2c} \geq 0$ and $y = \ln(K/S_0)$. Function $u^{(-)}$ has a similar form.

## 6. Quantile hedging

Consider an admissible strategy $\pi = (\varphi_t, \psi_t)$ with capital $F_t = F_t^\pi = \varphi_t S(t) + \psi_t B(t) \geq 0$ for all $t \in [0, T]$. For a given payoff function $f(S(T))$ and given admissible strategy $(\varphi_t, \psi_t)$ with the initial capital $v$, we define the set of successful hedging as

$$A = A(v, \varphi, \psi, f) = \{\omega : B(T)^{-1} F_T \geq f\}.$$

If a strategy is a perfect hedge, then $\mathsf{P}_\sigma(A(\mathfrak{c}^\sigma, \varphi, \psi, f)) = 1$, which requires the initial capital $\mathfrak{c}^\sigma = \mathsf{E}_\sigma^*[B(T)^{-1} f]$. The problem of quantile hedging is to maximize probability of $A$ under the budget restriction $v_0$:

$$\begin{cases} \mathsf{P}_\sigma(A(v, \varphi, \psi, f)) \to \max \\ v \leq v_0 < \mathsf{E}_\sigma^* \left(B(T)^{-1} f\right) = \mathfrak{c}^\sigma. \end{cases} \tag{60}$$

It is known (see Föllmer and Leukert [11]) that (60) is equivalent to the following optimization problem

$$\begin{cases} \mathsf{P}_\sigma(A) \to \max, \\ \mathsf{E}_\sigma^* \left(B(T)^{-1} f \cdot \mathbf{1}_A\right) \leq v_0. \end{cases} \tag{61}$$

Let $\tilde{A} = \tilde{A}_\sigma$ be the solution of (61). The perfect hedge $(\tilde{\varphi}, \tilde{\psi})$ with initial capital $v_0$ for the claim $\tilde{f} = f \cdot \mathbf{1}_{\tilde{A}}$ is the solution of (60) and its set of successful hedging $A = A(v, \tilde{\varphi}, f)$ coincides with $\tilde{A}$.

Moreover, the structure of the set $\tilde{A}$ is

$$\tilde{A} = \left\{ \frac{\mathrm{d}\mathsf{P}_\sigma}{\mathrm{d}\mathsf{P}_\sigma^*} \Big|_T \geq \gamma \cdot f \right\}, \quad \gamma = \text{const}, \ \gamma > 0. \tag{62}$$

Using (9)-(10) and (5), we obtain

$$\frac{\mathrm{d}\mathsf{P}_\sigma^*}{\mathrm{d}\mathsf{P}_\sigma} \Big|_T = \mathcal{E}_T(X_*^\sigma + J_*^\sigma) = \mathrm{e}^{X_*^\sigma(T)} \kappa_*^\sigma(T) = \mathrm{e}^{aX^\sigma(T) + bT} \kappa_*^\sigma(T),$$

where $a = \frac{c_+^* - c_-^*}{c_+ - c_-}$ and $b = \frac{c_+ c_-^* - c_- c_+^*}{c_+ - c_-}$. Hence, the set of successful hedging $\tilde{A}$ can be represented as

$$\tilde{A} = \tilde{A}^\gamma = \left\{ \mathrm{e}^{-aX^\sigma(T)} \geq \gamma \mathrm{e}^{bT} \kappa_*^\sigma(T) \cdot f \right\}.$$

For the standard call option with $f = (S(T) - K)^+ = \left(S_0 \mathrm{e}^{X^\sigma(T)} \kappa^\sigma(T) - K\right)^+$, the set $\tilde{A}$ has the form

$$\tilde{A} = \left\{ \mathrm{e}^{-aX^\sigma(T)} > \gamma \mathrm{e}^{bT} \kappa_*^\sigma(T) \left(S_0 \mathrm{e}^{X^\sigma(T)} \kappa^\sigma(T) - K\right)^+ \right\} = \bigsqcup_{n=0}^\infty A_n,$$

where

$$A_n = \left\{ \mathrm{e}^{-aX^\sigma(T)} \geq \gamma \kappa_{n,\sigma}^* \mathrm{e}^{bT} \left(S_0 \kappa_{n,\sigma} \mathrm{e}^{X^\sigma(T)} - K\right)^+, \ N^\sigma(T) = n \right\}.$$

In the case $-a \leq 1$, the sets $A_n$ take the form

$$A_n = \{X^\sigma(T) \leq y_n,\ N^\sigma(T) = n\}.$$

Here $y_n = y_n(\gamma) = \ln z_n - b_n^{(\sigma)}$, $b_n^{(\sigma)} = \ln \kappa_{n,\sigma}$, and $z_n = z_n(\gamma)$ is the unique solution of the algebraic equation

$$z^{-a} = \gamma \kappa_{n,\sigma}^* (\kappa_{n,\sigma})^{-a} e^{bT} (S_0 z - K)^+. \tag{63}$$

It is clear that $y_n = y_n(\gamma)$ decreases in $\gamma$ and $y_n \geq \ln K/S_0 - b_n^{(\sigma)}$.

To find the constant $\gamma$, we consider the equation

$$v_0 = S_0 \sum_{n=0}^{\infty} \left[ U_n^{(\sigma)}(y - b_n^{(\sigma)},\ T) - U_n^{(\sigma)}(y_n(\gamma),\ T) \right]$$

$$- K \sum_{n=0}^{\infty} \left[ u_n^{(\sigma)}(y - b_n^{(\sigma)},\ T) - u_n^{(\sigma)}(y_n(\gamma),\ T) \right], \tag{64}$$

where $u_n^{(\sigma)}$ and $U_n^{(\sigma)}$, $n \geq 0$, $\sigma = \pm 1$ are defined in (41)-(42), $y = \ln K/S_0$. As $v_0 < \mathfrak{c}^\sigma$, the equation has a unique solution $\gamma = \gamma(v_0)$ due to the monotonicity of $y_n = y_n(\gamma)$. Now the probability of maximal successful hedging set can be calculated as

$$\mathsf{P}(\tilde{A}) = \sum_{n=0}^{\infty} \mathsf{P}(A_n) = 1 - \sum_{n=0}^{\infty} u_n^{(\sigma)}(y_n(\gamma),\ T;\ \lambda_\pm,\ c_\pm,\ 0). \tag{65}$$

**Example 6.1** Let $\lambda_+ = \lambda_- = \frac{r_+ - c_+}{h_+} = \frac{r_- - c_-}{h_-}$. It means that the initial measure $\mathsf{P}$ is the martingale measure, and the corresponding process $N$ is a homogeneous Poisson process with such intensity. Hence $y_n \equiv \ln \frac{K + 1/\gamma}{S_0} - b_n^{(\sigma)}$, $a = b = 0$, and the equation (64) for $\gamma = \gamma(v_0)$ takes the form

$$v_0 = \mathfrak{c}(K,\ T) - \mathfrak{c}(K + 1/\gamma,\ T) - \frac{1}{\gamma} u^{(\sigma)}(K + 1/\gamma,\ T),$$

where $u^{(\sigma)}(z,\ T) = \sum_{n=0}^{\infty} u_n^{(\sigma)}(z - y_n^{(\sigma)},\ T)$, and $u_n^{(\sigma)}$, $n \geq 0$ are defined in (51), with $\lambda_\pm^* = \lambda_\pm$ and $r_\pm = 0$. The probability of successful hedging is equal to

$$\mathsf{P}(\tilde{A}) = \mathsf{P}_\sigma(S(T) < K + 1/\gamma)$$

$$= 1 - \sum_{n=0}^{\infty} u_n^{(\sigma)}\left( \ln \frac{K + 1/\gamma}{S_0},\ T;\ \lambda_\pm,\ c_\pm,\ 0 \right),\quad \gamma = \gamma(v_0).$$

Let $-a > 1$, then we have

$$A_n = \left\{ X^\sigma(T) \leq y_n^{(1)},\ N^\sigma(T) = n \right\} \bigcup \left\{ X^\sigma(T) \geq y_n^{(2)},\ N^\sigma(T) = n \right\}.$$

Here $y_n^{(1)} = \ln z_n^{(1)} - b_n^{(\sigma)}$ and $y_n^{(2)} = \ln z_n^{(2)} - b_n^{(\sigma)}$, where $z_n^{(1)}$ and $z_n^{(2)}$ are the solutions of (63).

The equation for $\gamma$ has the form

$$v = S_0 \sum_{n=0}^{\infty} \left[ U_n^{(\sigma)}(y - b_n^{(\sigma)}, \ T) - U_n^{(\sigma)}(y_n^{(1)}, \ T) + U_n^{(\sigma)}(y_n^{(2)}, \ T) \right]$$

$$-K \sum_{n=0}^{\infty} \left[ u_n^{(\sigma)}(y - b_n^{(\sigma)}, \ T) - u_n^{(\sigma)}(y_n^{(1)}, \ T) + u_n^{(\sigma)}(y_n^{(2)}, \ T) \right]$$

and the solution of the quantile hedging problem is

$$\mathsf{P}_\sigma(\tilde{A}) = 1 - \sum_{n=0}^{\infty} \left[ u_n^{(\sigma)}(y_n^{(1)}, \ T; \ \lambda_\pm, \ c_\pm, \ 0) - u_n^{(\sigma)}(y_n^{(2)}, \ T; \ \lambda_\pm, \ c_\pm, \ 0) \right]. \tag{66}$$

The dual problem

$$\begin{cases} v \to \min \\ \mathsf{P}_\sigma\left(A(v, \ \varphi, \ f)\right) \geq 1 - \varepsilon \end{cases} \tag{67}$$

minimizes the initial capital under a fixed risk level. It can be solved as follows. Using (65) and (66), we can find $\gamma$ from the equation $\mathsf{P}_\sigma(\tilde{A}^\gamma) = 1 - \varepsilon$, i. e.

$$\sum_{n=0}^{\infty} u_n^{(\sigma)}(y_n(\gamma), \ T; \ \lambda_\pm, \ c_\pm, \ 0) = \varepsilon \qquad (\text{for } -a \leq 1), \tag{68}$$

$$\sum_{n=0}^{\infty} \left[ u_n^{(\sigma)}(y_n^{(1)}(\gamma), \ T; \ \lambda_\pm, \ c_\pm, \ 0) - u_n^{(\sigma)}(y_n^{(2)}(\gamma), \ T; \ \lambda_\pm, \ c_\pm, \ 0) \right] = \varepsilon \tag{69}$$

$$(\text{for } -a > 1),$$

where $y_n = \ln z_n - b_n^{(\sigma)}$, and $z_n = z_n(\gamma)$, $n \geq 0$ solve the equation (63). The set of successful hedging $\tilde{A}$ is now defined and the optimal strategy is the perfect hedge of the claim $f \cdot \mathbf{1}_{\tilde{A}}$.

**Remark 1:** We would like to mention here a possible application of this type of hedging to risk-management of equity-linked life insurance contracts. The payoffs of such contracts depend on the evolution of risky assets as well as a survival status of insureds during the contract period $[0, T]$. Denote $T(x)$ the remaining lifetime of a policy holder, who is currently of age $x$. Then the future payment may have the form $\max\{S(T), \ K\} I_{\{T(x) > T\}}$, where $K$ is the so-called maturity guarantee. It is natural to assume that $T(x)$ does not depend on the stock market. Next, rewriting $\max\{S(T), \ K\} = K + (S(T) - K)^+$, we reduce the pricing problem to that of a related call option. If $_T p_x = \mathsf{P}(T(x) > T)$ is the survival probability, we can find the Brennan and Schwartz [5] price

$$_T c_x = \mathsf{E}_\sigma^* \left[ B(T)^{-1} f \cdot \mathbf{1}_{\{T(x) > T\}} \right] =_T p_x \cdot \mathsf{E}_\sigma^* \left[ B(T)^{-1} f \right], \tag{70}$$

widely exploited in this area. This price is smaller than the fair price for the call option. Hence, perfect hedging is impossible with the help of underlying assets in the market. On the other hand, we can consider $_T c_x$ as $v_0$ in the context of quantile hedging above. For given $T$ and $x$ parameter $_T p_x$ can be found from actuarial life tables. Taking $v_0 =_T c_x$, we can construct the set $\tilde{A}$ and the corresponding quantile hedge according to (60)-(61).

**Remark 2:** In the framework of the model (18)-(19) it is possible to analyze optimal investment problems [16], as well as shortfall risk minimization [12]. This analysis will be reported elsewhere later. Seeking for simplicity we omit these results here.